\documentclass[aps,prl,article,twocolumn,showpacs,preprintnumbers,amsmath,amssymb,superscriptaddress]{revtex4}

\date{\today}
\usepackage{epsfig}
\usepackage{subfigure}
\usepackage{graphicx}
\usepackage{dcolumn}
\usepackage{bm}
\usepackage[colorlinks,linkcolor=blue,hyperindex,CJKbookmarks]{hyperref}
\usepackage{float}
\usepackage{hyperref}
\usepackage{comment}
\newcommand{\Ham}   {{\mathcal{H}}}
\newcommand{\kbf}      {\textbf{k}}
\newcommand{\qbf}      {\textbf{q}}

\newcommand{\ibf}      {\textbf{i}}
\newcommand{\jbf}      {\textbf{j}}

\begin{document}

\title{Magnon Splitting Induced by Charge Transfer in the Three-Orbital Hubbard Model}
\author{Yao Wang }
 \affiliation{SLAC National Accelerator Laboratory, Stanford Institute for Materials and Energy Sciences, 2575 Sand Hill Road,
Menlo Park, California 94025, USA}
 \affiliation{Department of Applied Physics, Stanford University, California 94305, USA}
  \affiliation{Department of Physics, Harvard University, Cambridge, Massachusetts 02138, USA}
\author{Edwin W. Huang}%
\affiliation{SLAC National Accelerator Laboratory, Stanford Institute for Materials and Energy Sciences, 2575 Sand Hill Road,
Menlo Park, California 94025, USA}
\affiliation{Department of Physics, Stanford University, California 94305, USA}%
\author{Brian Moritz}
\affiliation{SLAC National Accelerator Laboratory, Stanford Institute for Materials and Energy Sciences, 2575 Sand Hill Road,
Menlo Park, California 94025, USA}%
\affiliation{Department of Physics and Astrophysics, University of North Dakota, Grand Forks, North Dakota 58202, USA}
\author{Thomas P. Devereaux}
 \email[Author to whom correspondence should be addressed to Y. W. (\href{mailto:yaowang@g.harvard.edu}{yaowang@g.harvard.edu}) or T.P. D. (\href{mailto:tpd@stanford.edu}{tpd@stanford.edu})
]{}
\affiliation{SLAC National Accelerator Laboratory, Stanford Institute for Materials and Energy Sciences, 2575 Sand Hill Road,
Menlo Park, California 94025, USA}%
\affiliation{Geballe Laboratory for Advanced Materials, Stanford University, California 94305, USA}
\date{\today}
\begin{abstract}
Understanding spin excitations and their connection to unconventional superconductivity have remained central issues since the discovery of the cuprates. Direct measurement of the dynamical spin structure factor in the parent compounds can provide key information on important interactions relevant in the doped regime, and variations in the magnon dispersion have been linked closely to differences in crystal structure between families of cuprate compounds. Here, we elucidate the relationship between spin excitations and various controlling factors thought to be significant in high-$T_c$ materials by systematically evaluating the dynamical spin structure factor for the three-orbital Hubbard model, revealing differences in the spin dispersion along the Brillouin zone axis and the diagonal. Generally, we find that the absolute energy scale and momentum dependence of the excitations primarily are sensitive to the effective charge-transfer energy, while changes in the on-site Coulomb interactions have little effect on the details of the dispersion. In particular, our result highlights the splitting between spin excitations along the axial and diagonal directions in the Brillouin zone.  This splitting decreases with increasing charge-transfer energy and correlates with changes in the apical oxygen position, and general structural variations, for different cuprate families.
\end{abstract}
\maketitle

\newcommand{\Jeff}{$J_{\rm eff}$}
\newcommand{\Jdiff}{$\Delta E_{\rm AFZB}$}
\newcommand{\EmagTop}{$E_{\rm magnon}$}
\newcommand{\ctGap}{$E_{CT}$}
Layered copper-oxide materials, or cuprates, have been the focus of numerous studies since the discovery of high-$T_c$ superconductivity. In contrast to the debate over the origins of unconventional superconductivity, the pseudogap, and novel charge ordered states which emerge upon doping, the ground state of undoped cuprates has been demonstrated unequivocally to be a Mott (or more precisely charge-transfer) insulator with long-range antiferromagnetic order.\cite{kivelson2003detect,keimer2015quantum}. While a number of physical properties in the parent compounds correlate with the maximum superconducting transition temperature $T_c^{\rm max}$, it generally has been believed that decreasing the charge-transfer energy $\Delta$, hence the parent compound's charge-transfer gap, boosts $T_c^{\rm max}$\cite{weber2012scaling, ruan2016relationship}. 

Factors including the kinetic energy\cite{ohta1991charge, weber2012scaling}, oxygen hole density\cite{tallon1990therelationship, rybicki2016perspective}, and the importance of apical oxygen ligands\cite{ohta1990transition, ohta1991charge, ohta1991apex, weber2012scaling} have been implicated directly or indirectly in the variation of $T_c^{\rm max}$ between different cuprate families through the charge-transfer energy. At the same time, spin fluctuations have been deemed a leading candidate for the pairing glue in unconventional superconductivity\cite{scalapino2012common}, with significant attention given to the correlation between the spin excitation spectrum and $T_c^{\rm max}$.  While the superexchange energy \Jeff\ in a simple perturbative analysis decreases with increasing $\Delta$\cite{ohta1991charge, ruan2016relationship}, the relationship between the magnon dispersion in the parent compounds, the evolution of the spin excitation spectrum with doping, and ultimately $T_c^{\rm max}$ remains under debate\cite{munoz2000accurate, mallett2013dielectic, wulferding2014relation, tallon2014anomalous, ellis2015correlation}.

With recent improvements in energy resolution, resonant inelastic x-ray scattering (RIXS) has emerged as a powerful tool complementing neutron scattering for studying the detailed spin excitation spectrum in materials\cite{braicovich2009dispersion, le2011intense, schlappa2012spin, dean2013persistence,dean2013high, lee2014asymmetry, ishii2014high, jia2014persistent,jia2016using, ament2010strong}. Recently, RIXS experiments were performed in a number of cuprate compounds using polarization control in a cross-polarized geometry\cite{peng2017influence}.  These measurements provided a new perspective on the relationship between superconductivity and the spin degrees of freedom, particularly in the variation of the spin excitations along the antiferromagnetic zone boundary (AFZB) between different parent compounds. The momentum-resolved magnon dispersion -- its energy, width, and intensity -- reflects a rich set of information, in contrast to the scalar superexchange energy \Jeff\ in a Heisenberg-like picture for the spin degrees of freedom in the cuprates. These extra information can be useful in exploring a more detailed link between spin excitations and the emergent physics of high-$T_c$ superconductivity.%, beyond simple \Jeff\ modeling.

For this purpose, we have performed a systematic study of the dynamical spin structure factor in the three-orbital Hubbard model using both exact diagonalization (ED) and determinant quantum Monte Carlo (DQMC) approaches.  This model allows us to capture the explicit dependence of the spin excitations on realistic oxygen degrees of freedom, which can reflect differences in the structural and chemical environment between different cuprate families.  We find magnons along both the nodal and antinodal directions to be insensitive to the on-site Coulomb interactions; however, the magnon energy, width, and intensity intimately depend on the kinetic and charge-transfer energies.  Unlike the behavior in the linear spin-wave or Heisenberg pictures, the magnon energy displays a non-monotonic dependence on these factors.  At the same time, the effective charge-transfer energy controls a monotonic nodal-antinodal splitting, which hints at the connections between the structure, the charge-transfer energy, this magnon splitting, and ultimately $T_c^{\rm max}$.

%%%%%%%%%%%%% model %%%%%%%%%%%%%%%%%%%%%%%%

%------------- Figure 1 Begin -------------%
\begin{figure}[!t]
\begin{center}
\includegraphics[width=\columnwidth]{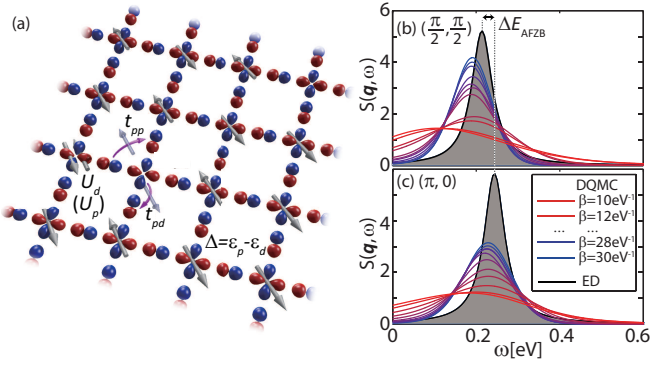}
\caption{\label{fig:1} (a) A sketch of the three-orbital model for cuprates with blue and red colors denoting positive and negative lobes of the orbital wavefunctions ($d_{x^2-y^2}$ for copper and $p_{x/y}$ for oxygen). Arrows indicate the kinetic energy, or hopping, terms $t_{pd}$ and $t_{pp}$.  (b,c) The spin structure factor along the (b) nodal and (c) antinodal directions calculated using DQMC at various temperatures (colored lines from red to blue) and ED at zero temperatures $T=1/\beta$ (shaded black curves). An \textit{ad hoc} Lorentzian broadening (HWHM) $\Gamma=30$meV has been adopted in the ED calculations to approximate final state lifetime effects. The gray dashed lines denote the magnon energies \EmagTop, while the black arrow defines the magnon energy splitting \Jdiff.
}
\end{center}
\end{figure}
%------------- Figure 1 End -------------% 

The electronic and magnetic properties of cuprates can be captured in minimal copper-oxide models representing the copper $3d_{x^2-y^2}$ and oxygen $2p_{x/y}$ degrees of freedom\cite{kung2016characterizing}. The Hamiltonian for the three-orbital Hubbard model\cite{mattheiss1987electronic, varma1987charge, emery1987theory}, whose terms are sketched in Fig.~\ref{fig:1}(a), reads as
\begin{eqnarray}
\mathcal{H}&=&\left[-\sum_{\langle \ibf,\jbf\rangle \sigma}t_{pd}^{\ibf\jbf} d^{\dagger}_{\ibf\sigma}p_{\jbf\sigma}-\sum_{\langle \jbf,\jbf'\rangle \sigma}t_{pp}^{\jbf\jbf'}p^{\dagger}_{\jbf,x\sigma}p_{\jbf',y\sigma}\right]+h.c.\nonumber\\
&&+\Delta \sum_{\jbf\sigma}n^p_{\jbf\sigma}+U_d\sum_\ibf n^d_{\ibf\uparrow}n^d_{\ibf\downarrow}+U_p\sum_{\jbf\alpha} n^p_{\jbf\uparrow}n^p_{\jbf\downarrow}
\end{eqnarray}
where $d^\dagger_{\ibf\sigma}$ ($p^\dagger_{\jbf\sigma}$) creates a hole with spin $\sigma$ at a copper site $\ibf$ (oxygen site $\jbf$) and $n^d_{\ibf\sigma}$ ($n^p_{\jbf\sigma}$) is the copper (oxygen) hole number operator. The first two terms of the Hamiltonian represent the nearest-neighbor copper-oxygen and oxygen-oxygen hybridization. The hopping integrals ($t_{pd}^{\ibf\jbf}$ and $t_{pp}^{\jbf\jbf'}$) change sign with a $d$-like symmetry. The third term in the Hamiltonian represents the charge-transfer energy $\Delta=\varepsilon_p-\varepsilon_d$, while the last two terms are the on-site Hubbard interactions for the copper and oxygen, respectively. We use the following baseline set of parameters (in units of eV): $\Delta=3.23$; $U_d =8.5$, $U_p =4.1$; $|t_{pd}|=1.13$, $|t_{pp}| = 0.49$ that have yielded satisfactory fits to angle resolved photoemission as well as oxygen and copper x-ray spectroscopic data \cite{chen2013doping, katagiri2011theory, andersen1994plane, hybertsen1989calculation, andersen1995lda}.

Spin excitations and fluctuations are reflected by the dynamical spin structure factor
\begin{equation}
S(\textbf{q},\omega)=\frac1{\pi}\textrm{Im}\Big\langle G\Big|\rho_{-\textbf{q}}\frac1{\Ham-E_G-\omega-i\Gamma}\rho_{\textbf{q}}\Big|G \Big\rangle,
\end{equation}
Here, the spin density operator $\rho_{\textbf{q}}=\sum_{\textbf{k}\sigma}\sigma \psi_{\kbf+\qbf,\sigma}^\dagger \psi_{\kbf\sigma} $ with $\psi_{\kbf\sigma} = \left( d_{\kbf\sigma}, p_{\kbf,x\sigma}, p_{\kbf,y\sigma}\right)^T$, where $|G\rangle$ and $E_G$ are the ground state wavefunction and energy, respectively.

Unlike the linear spin-wave theory, applicable to simpler spin models, predicting the spin excitations for the three-orbital model represents a highly nontrivial task due to the additional degrees of freedom.  For small cluster ED, we use the parallel Arnoldi method\cite{lehoucq1998arpack} to determine the ground state wavefunction and the continued fraction expansion\cite{dagotto1994correlated} to obtain $S({\bf q},\omega)$. For the DQMC simulations, the imaginary time correlators are analytically continued into real frequency spectra through the Maximum Entropy method \cite{bryan1990maximum, jarrell1996bayesian}.  As described in Ref.~\onlinecite{jarrell1996bayesian}, we use model functions based on the first moments of the spectra. Other model functions such as using spectra from a higher temperature simulations produce similar spectra with negligible differences ($<10$meV) in peak positions. Unless noted, calculations were performed on 8-site Betts clusters\cite{betts1999improved} with periodic boundary conditions, which represents a minimal cluster that contains both the nodal and antinodal momenta at issue in the study, although finite-size effects remain. 

Figs.~\ref{fig:1}(b) and (c) show $S(\textbf{q},\omega)$ obtained using ED at zero temperature.  These are supplemented by results from DQMC calculations at various temperatures. The asymptotic consistency between these two techniques confirms the existence of a substantial difference in magnon energy \Jdiff\ between the nodal and antinodal directions. Unlike the phenomenological fitting with complex Heisenberg models in Ref.~\onlinecite{peng2017influence}, we find such an asymmetry can be naturally illustrated by taking into account the charge-transfer nature of cuprate microscopic models.

%------------- Figure 2 Begin -------------%
\begin{figure}[!t]
\begin{center}
\includegraphics[width=\columnwidth]{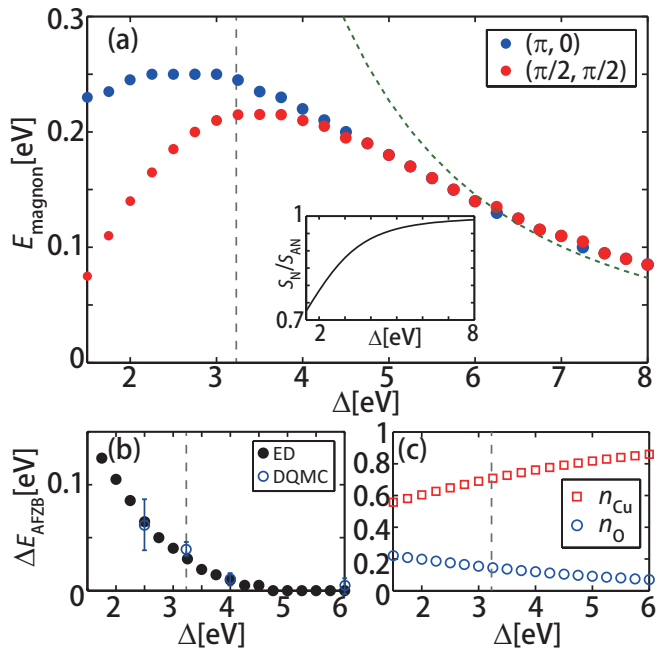}
\caption{\label{fig:2} (a) Nodal (red) and antinodal (blue) spin excitations as a function of charge-transfer energy $\Delta$. The solid dots represent the peak energies of $S(\qbf,\omega)$, while the size of each dot corresponds to the integrated intensity of $S(\qbf,\omega)$. The green dashed line in (a) marks the energy of the top of the magnon band as predicted by linear spin-wave theory using a perturbative $J_{\rm eff}$. The insets show the ratio between the nodal and antinodal intensities. 
(b) Absolute magnon energy splitting $\Delta E_{\rm AFZB}$ and (c) average hole occupancy $n_{\rm Cu}$ on copper (red squares) and $n_{\rm O}$ on oxygen (blue circles) as a function of $\Delta$. The grey dashed lines indicate the canonical $\Delta$, as noted in the main text.
The errorbars in (b) for DQMC data represent random errors determined by bootstrap resampling.
}
\end{center}
\end{figure}
%------------- Figure 2 End -------------%

The nodal and antinodal magnon energies \EmagTop\ are extracted from the positions of the peaks in the spin excitation spectrum as shown in Fig.~\ref{fig:1}(b).  Figure~\ref{fig:2}(a) clearly shows that these energies are tuned through the effective charge-transfer $\Delta$.  As a comparison in the limit where the kinetic energy scale is much smaller than the charge-transfer ($t_{pd}\ll \Delta$), the effective (nearest-neighbor) spin exchange energy is given by $J_{\rm eff}\approx \frac{4t_{pd}^4}{\Delta^2}\left(\frac1{\Delta} + \frac2{U_d}\right)$\cite{anderson1959spin, zhang1988effective, muller2002derivation,  zaanen1988canonical}. Such a Heisenberg picture leads to linear spin-waves with maximal energies that vary as a function of $\Delta$ as shown by the green dashed line.  In the large $\Delta$ limit, the oxygen degrees of freedom become irrelevant  and the magnons display little if any variation in energy splitting or intensity along the AFZB.

In the opposite limit of small $\Delta$, holes in the parent compound move to the oxygen sublattice.  This decreases the influence of electronic interactions by reducing copper double occupancies, and in turn lowers the effective spin excitation energy along the nodal direction, where the oxygen spins initially align.  In contrast to the limit of large $\Delta$, the magnon energies along the nodal and antinodal directions split, which reflects an over-estimation of the effective kinetic energy in the perturbative expansion for \Jeff\ as $\Delta$ decreases and the extra holes hinder copper-oxygen and intra-oxygen hole motion. These holes also dilute the net magnetic moment on the oxygen orbitals, reflected by the rapid drop in the spectral weight ratio $S_{\rm N}/S_{\rm AN}$ [see the inset of Fig.~\ref{fig:2}(a)].  The simple spin-wave estimate \Jeff\ fails to predict details of the magnon energies and weights, even in the regime of the canonical charge-transfer energy ($\sim 3$eV).

While the magnon energies are non-monotonic as a function of $\Delta$, the magnon splitting \Jdiff, defined as $E_{\rm magnon}(\pi,0)-E_{\rm magnon}(\pi/2,\pi/2)$, always decreases with increasing $\Delta$ [see Fig.~\ref{fig:2}(b)]. As $\Delta$ increases and the holes are redistributed [see Fig.~\ref{fig:2}(c)], carriers become localized to copper and the energy for both nodal and antinodal spin excitations merge, a signature of Heisenberg physics. Such a negative correlation between $\Delta$, $n_{\rm O}$, and \Jdiff\ is consistent with previous experiments that also link $\Delta$ to $T_c^{\rm max}$\cite{weber2012scaling, ruan2016relationship, tallon1990therelationship, rybicki2016perspective}.  Additionally, we compare the relative magnon splitting evaluated using ED with that obtained from DQMC calculations for two different sizes and temperatures. While high temperatures may overestimate this splitting, the trend is consistent with the ED calculation, further supporting the existence of this inverse relationship in cuprates [see the Supplementary Materials in Ref.~\onlinecite{SUPMAT}].
 
%------------- Figure 4 Begin -------------%
\begin{figure}[!t]
\begin{center}
\includegraphics[width=\columnwidth]{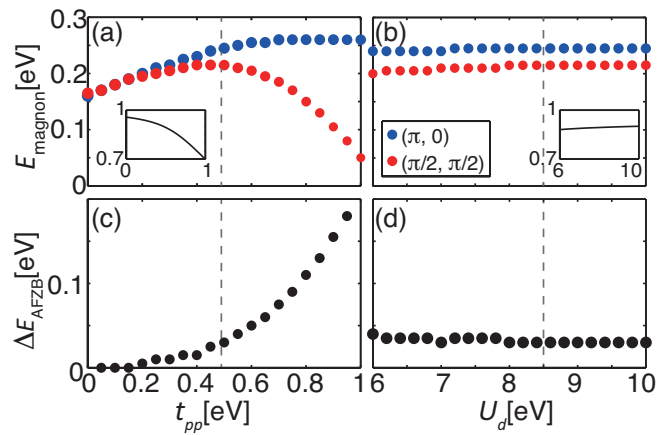}
\caption{\label{fig:4} (a,b) Nodal (red) and antinodal (blue) spin excitations as a function of (a) oxygen hopping $t_{pp}$ and (b) on-site Columb repulsion $U_d$ on copper following the conventions of Fig.~\ref{fig:2}(a). The insets show the ratio of nodal to antinodal intensity. (c,d) The magnon splitting \Jdiff\ as a function of (c) $t_{pp}$ and (d) $U_d$. The grey dashed lines indicate the canonical parameter values.  
}
\end{center}
\end{figure}
%------------- Figure 4 End -------------%

%\subsection{Kinetic Energy}
Apart from the charge-transfer energy, the intra-oxygen hopping, as parameterized by $t_{pp}$, competes with copper-oxygen hopping to affect the band structure and strongly impact the spin structure factor, and even potentially $T_c$\cite{ohta1991charge, weber2012scaling}. 
As shown in Fig.~\ref{fig:4}(a), small $t_{pp}$ leads to asymptotically similar nodal and antinodal spin excitations, since the three-orbital model can be re-expressed as an effective single-band model in this limit.  As $t_{pp}$ increases, approaching $t_{pd}$, the oxygens form a new sublattice and the relatively large $n_{\rm Cu}$ and spin density on copper, set by the charge-transfer energy, lead to a reduced $S_N$.  Equivalently, the increased oxygen kinetic energy reduces the effective oxygen site energy, and by extension the effective charge-transfer energy $\Delta$ felt by the carriers. Thus, similar trends come from increasing $t_{pp}$ or reducing $\Delta$ [see Fig.~\ref{fig:4}(c)].

This modulation of spin excitations by kinetic energy would be influenced by tuning yet longer-range copper-copper, copper-oxygen, and oxygen-oxygen hoppings.
For example, the nearest-neighbor copper [see the Supplementary Materials \cite{SUPMAT}] and next-nearest-neighbor oxygen hopping $t_{pp^\prime}$ \cite{weber2012scaling} can be used to tune an effective $\Delta$, which for the scope of this work will be sufficient to underscore the important controlling parameters in the model.

In charge-transfer insulators like cuprates, the energy gap is determined primarily by the Madelung potentials, electron affinities, and ionization energies, rather than the on-site, screened Coulomb repulsion. Here we extend this conclusion to the spin excitations. 
Our numerical simulations reveal that over a wide range of both $U_d$ [see Fig.~\ref{fig:4}(b)] and $U_p$ [see the Supplementary Materials \cite{SUPMAT}], the spin excitation energies change little, if at all, compared to their dependence on the effective charge-transfer energy and kinetic energy [see Fig.~\ref{fig:2}(a) and Fig.~\ref{fig:4}(a)].
This can be understood by considering once again the distribution of holes in the CuO$_2$ planes: provided that $U_d>\Delta$, the low-energy physics including that of the spin excitations, depends on the singly-occupied copper and oxygen states. Thus in the atomic limit, the controlling parameter is the charge-transfer energy. 
The small dependence on Coulomb repulsion is consistent with the linear spin wave prediction with a perturbative $J_{\rm eff}$ in the physically relevant range of parameters: varying $U_d$ from 6 to 8\,eV modulates the magnon energy in a range of $0.45-0.49eV$. 

%\section{Discussions and Conclusion}
%------------- Figure 6 Begin -------------%
\begin{figure}[!t]
\begin{center}
\includegraphics[width=\columnwidth]{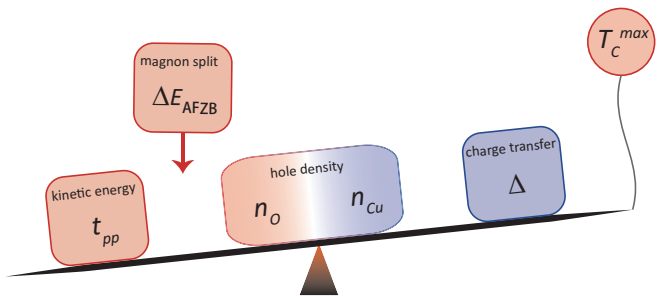}
\caption{\label{fig:6} Summary of factors that affects $T_c^{\rm max}$. The red boxes (kinetic energies, hole density on oxygens and magnon splitting) indicate enhancing factors, while the blue boxes (charge-transfer energy $\Delta$ and hole density on coppers) indicate suppressing factors.
}
\end{center}
\end{figure}
%------------- Figure 6 End -------------%

As discussed extensively over the past thirty years, the maximum superconducting transition temperature has been found to be (positively or negatively) correlated with many factors [see Fig.~\ref{fig:6}]. 
These dependencies originate from differences in the charge-transfer energy and distribution of carriers in electronic structure theory\cite{weber2012scaling, ruan2016relationship, ohta1991charge, tallon1990therelationship, rybicki2016perspective}, or correspondingly the separation from the apical ligands in the crystal structure\cite{ohta1990transition, ohta1991apex}.
Here we add an important building block to our understanding of these cross-correlations, showing that the magnon energies in a three-orbital model do not follow the monotonic relationship predicted by perturbation theory ($\sim$\Jeff).  Instead, the magnon splitting between nodal and antinodal directions \Jdiff\ displays a strong monotonic dependence on the effective charge-transfer energy, consistent with the positive correlation between \Jdiff\ and $T_c^{\rm max}$ proposed by recent experiments\cite{peng2017influence}.

To summarize, variations in the spin dispersion measured for different cuprate families can be attributed to differences in structure and crystal field effects rather than to differences in on-site Coulomb repulsion.  We have shown that the three-orbital model can naturally capture the asymmetry in the spin response between the nodal and antinodal directions observed in experiments, without introducing phenomenological parameters for long-range spin exchange\cite{peng2017influence}.
In contrast to the average magnon energy or perturbative spin exchange energy, the monotonic modulation of the nodal/antinodal magnon splitting by the charge-transfer energy represents a clear and robust trend. 
With help from RIXS experiments, such a correlation provides a better platform for discussions of the link between superconductivity and spin correlations for further investigations of high-$T_c$ materials.

\section{Acknowledgements}
We thank  L. Braicovich, G. Ghiringhelli and Y.Y. Peng for sharing the experimental data and insightful discussions. This material is based upon work supported by the U.S. Department of Energy, Office of Science, Office of Basic Energy Sciences, Division of Materials Sciences and Engineering, under Contract No. DE-AC02-76SF00515. Y.W. was supported by the Stanford Graduate Fellowship in Science and Engineering. This research used resources of the National Energy Research Scientific Computing Center (NERSC), a U.S. Department of Energy Office of Science User Facility operated under Contract No. DE-AC02-05CH11231.

\bibliography{paper}

\end{document}